\documentstyle[prb,aps,multicol]{revtex}
\begin{document}
\draft
\title{A Path integral approach to the scattering theory of quantum transport} 
\author{D. Endesfelder\cite{present_address}}
\address{Oxford University,
Theoretical Physics, \\
1 Keble Road, \\
United Kingdom}
\date{\today}
\maketitle
\begin{abstract}
The scattering theory of quantum transport relates transport properties 
of disordered mesoscopic conductors to their transfer matrix $\bbox{T}$. 
We introduce a novel approach to the statistics of transport quantities
which expresses the probability distribution of $\bbox{T}$ as a path 
integral. The path integal is derived for a model of conductors with 
broken time reversal invariance in arbitrary dimensions. It is applied 
to the Dorokhov-Mello-Pereyra-Kumar (DMPK) equation which describes 
quasi-one-dimensional wires. We use the equivalent channel model whose 
probability distribution for the eigenvalues of $\bbox{TT}^{\dagger}$ is 
equivalent to the DMPK equation independent of the values of the forward 
scattering mean free paths. We find that infinitely strong forward 
scattering corresponds to diffusion on the coset space of the transfer 
matrix group. It is shown that the saddle point of the path integral 
corresponds to ballistic conductors with large conductances. We solve the 
saddle point equation and recover random matrix theory from the saddle 
point approximation to the path integral.  
\end{abstract}
\begin{multicols}{2}
\section{Introduction}
Advances in microfabrication technology led to the realization of 
mesoscopic electronic devices. In such devices the mean free path for 
inelastic electron scattering exceeds the dimension of the device. As
a consequence phase coherence is maintained which leads to quantum 
interference effects like universal conductance fluctuations, persistent 
currents and Aharanov-Bohm oscillations in rings, or weak localization
\cite{MesoscopicRev}. The phase coherence has also serious theoretical
implications.  It causes large conductance fluctuations which are 
related to the problem of high gradient operators in the field theoretic 
description of the metal insulator transition \cite{Kravtsov85a,%
Altshuler86a,Kravtsov89a,Wegner90a}. These fluctuations manifest 
themselves already in the metallic regime as logarithmic normal tails
of the conductance probability distribution. As the critical regime 
is approached the conductance probability distribution becomes 
increasingly broader until it reaches a logarithmic normal
form in the insulating regime \cite{Altshuler91a}. 

A common approach to transport quantities of mesoscopic conductors 
is the scattering theory of quantum transport \cite{Landauer70a,%
Buettiker85a}. It models the conductor by a disordered region which is
connected to a number of ideal leads which support propagating wave modes. 
The number of leads corresponds to the number of measurement terminals. 
Here only two terminal geometries will be considered. The scattering 
matrix relates the amplitudes $I_{k}, I'_{k}$ of the incoming with the 
amplitudes $O_{k}, O'_{k}$ $(k=1,\ldots,N)$ of the scattered propagating
wave modes at the Fermi energy, 
\begin{equation}
\left(\begin{array}{c}
      O\\
      O'
      \end{array}\right)= 
\bbox{S}\left(\begin{array}{c}
              I\\
              I'
              \end{array}\right),
\end{equation}
where 
\begin{equation}
\bbox{S}=\left(\begin{array}{cc}
               \bbox{r} & \bbox{t}' \\
               \bbox{t} & \bbox{r}' 
               \end{array}\right),
\end{equation}
$\bbox{t}$ and $\bbox{r}$ are the transmission and reflection matrices
for incident waves from the left, and $\bbox{t}'$ and $\bbox{r}'$ are 
the transmission and reflection matrices for incident waves from the 
right. The dimensionless two-probe conductance $g=G/(e^{2}/h)$ in terms
of the transmission eigenvalues $T_{k}$ of $\bbox{tt}^{\dagger}$ is
\begin{equation}
g=\sum_{k=1}^{N} T_{k}.                         
\end{equation}
There are three universality classes which correspond to different 
physical situations. Conductors with time reversal invariance lie in 
the orthogonal universality class. The unitary universality class 
corresponds to conductors in which the time reversal symmetry is 
broken, e.g. by a magnetic field. Conductors with spin-flip scattering
processes but no time reversal symmetry breaking fall into the 
symplectic universality class. 

Recently the quasi-one-dimensional wire has attracted considerable 
attention. The width of a quasi-one-dimensional wire is of the order of 
the mean free path for elastic electron scattering so that transverse 
diffusion can be neglected and the cross section of the wire becomes 
structureless. Interesting non-perturbative results which are valid
for all wire lengths have been obtained for this system 
\cite{Zirnbauer92a,Mirlin94a,Frahm95a,Rejaei96a}. Furthermore it 
has been the ideal playground for new ideas in the field of quantum 
transport. 

One of these ideas is the Fokker-Planck (FP) approach to 
quasi-one-dimensional wires. The FP equation which describes the 
probability distribution for the transmission eigenvalues is known as the 
Dorokhov-Mello-Pereyra-Kumar (DMPK) equation. It has been derived by a 
number of authors \cite{Dorokhov82a,Mello88c,Mello91b,Macedo92a,Mello92a} 
who started from various different models. Its form is         
\begin{equation}
\frac{\partial p(s;\{\Gamma_{k}\})}{\partial s}=
\frac{2}{\gamma}\sum_{k=1}^{N}\frac{\partial}{\partial\Gamma_{k}}
\left(\frac{\partial p}{\partial\Gamma_{k}}+\beta p\frac{\partial
\Omega(\{\Gamma_{k}\})}{\partial\Gamma_{k}}\right),
\end{equation}
where 
\begin{eqnarray}
\Omega(\{\Gamma_{k}\}) & = & -\sum_{k<l}\ln |(\cosh\Gamma_{k}-
\cosh\Gamma_{l})/2| \nonumber\\
& & -1/\beta\sum_{k}\ln |\sinh\Gamma_{k}|,
\end{eqnarray} 
$\gamma=\beta N+2-\beta$, and $\cosh\Gamma_{k}=(2-T_{k})/T_{k}$. The 
values of $\beta$ are $1,2,$ and $4$ for the orthogonal, unitary and 
symplectic universality class respectively. The DMPK equation has been 
studied intensively in the past few years 
\cite{Mello88a,Mello88d,Hueffmann90a,Chalker93c,Macedo94b,Macedo94c,%
Macedo94a,Beenakker94b,Caselle95a,Brouwer96a}. Beenakker and Rejaei 
\cite{Beenakker93b,Beenakker94a} discovered that the variation 
\begin{equation}
p(s;\{\Gamma_{k}\})=\exp\left\{-\frac{\beta}{2}\Omega(\{\Gamma_{k}\})
\right\}\psi(s;\{\Gamma_{k}\})  
\end{equation}
of the Sutherland transformation \cite{Sutherland72a} which is known to 
solve the Brownian motion model for the circular unitary ensemble 
\cite{Pandey91a}, works as well for the DMPK equation. 
After this transformation $\psi(s;\{\Gamma_{k}\})$ obeys a Schr\"odinger 
equation for $N$ non-interacting particles. As a consequence the exact 
form of $p(s;\{\Gamma_{k}\})$ could be determined. This solution has been 
the basis for Frahm's exact calculation of the  one- and two-point 
correlation functions of the transmission eigenvalues \cite{Frahm95a}.

In this paper we present a novel approach to the scattering theory of
quantum transport which expresses the probability distribution of the 
transfer matrix as a path integral. Our motivation has been the belief 
that the path integral technique can be developped into a tool which is 
more powerful than the FP approach when it comes to the description of 
higher-dimensional conductors.   
\section{Scattering model}
We use the transfer matrix $\bbox{T}$ instead of the $\bbox{S}$-matrix to
model the scattering properties of the disordered conductor. The transfer 
matrix relates the scattering amplitudes in the left lead with the 
scattering amplitudes in the right lead 
\begin{equation}
\left(\begin{array}{c}
      O'\\
      I'
      \end{array}\right)= 
\bbox{T}\left(\begin{array}{c}
              I\\
              O
              \end{array}\right).
\end{equation}
It has the advantage that it obeys the multiplication law 
\begin{equation}
\bbox{T}(L+\delta L,0)=\bbox{T}(L+\delta L,L)\bbox{T}(L,0)
\end{equation}
which leads to the simple Langevin equation
\begin{eqnarray}
\label{Lvgl1}
\dot{\bbox{T}}(x)\equiv \frac{d\bbox{T}(x,0)}{dx} & = & 
\bbox{\varepsilon}(x)\bbox{T}(x,0) \nonumber \\
& \equiv & \left(\begin{array}{cc}
           \bbox{\varepsilon}^{11}(x) & \bbox{\varepsilon}^{12}(x) \\
           \bbox{\varepsilon}^{21}(x) & \bbox{\varepsilon}^{22}(x)
           \end{array}\right)\bbox{T}(x,0)
\end{eqnarray}
for the stochastic evolution of the transfer matrix. The disorder is 
generated by the multiplicative noise $\bbox{\varepsilon}$.  

In this paper we consider only conductors in the unitary universality 
class. Then, $\bbox{T}$ obeys the symmetry constraint
\begin{equation}
\label{Symm1}
\bbox{\Sigma}_{z}\bbox{T}^{\dagger}\bbox{\Sigma}_{z}\bbox{T}=\bbox{1}
\end{equation}
which ensures flux conservation, where
\begin{equation}
\bbox{\Sigma}_{z} = \left(\begin{array}{cc}
                          \bbox{1} & \bbox{0} \\
                          \bbox{0} & -\bbox{1}
                          \end{array}\right).
\end{equation}
A convenient parametrization of the transfer matrix is the polar 
decomposition \cite{Mello91b,Mello91a}
\begin{equation}
\label{PolDecomp}
\bbox{T}=\left(\begin{array}{cc} \bbox{u}_{1} & \bbox{0} \\
                                 \bbox{0} & \bbox{u}_{3}     
               \end{array}\right)
\left(\begin{array}{cc} 
      \cosh (\bbox{\Gamma}/\bbox{2}) & \sinh (\bbox{\Gamma}/\bbox{2}) \\
      \sinh (\bbox{\Gamma}/\bbox{2}) & \cosh (\bbox{\Gamma}/\bbox{2}) 
      \end{array}\right)
\left(\begin{array}{cc} \bbox{u}_{2} & \bbox{0} \\
                        \bbox{0} & \bbox{u}_{4} \end{array}\right),  
\end{equation}
where $\bbox{\Gamma}$ is a real, diagonal $N\times N$ matrix and 
$\bbox{u}_{i}$ $(i=1,2,3$ and $4)$ are unitary $N\times N$ matrices.  

The relation (\ref{Symm1}) implies that $\bbox{\Sigma}_{z}\bbox{
\varepsilon}^{\dagger}\bbox{\Sigma}_{z}+\bbox{\varepsilon}=0$ leading
to the symmetries   
\begin{eqnarray}
\label{Symm2}
\bbox{\varepsilon}^{11\:\dagger} & = & -\bbox{\varepsilon}^{11},\nonumber\\
\bbox{\varepsilon}^{22\:\dagger} & = & -\bbox{\varepsilon}^{22},\nonumber\\
\bbox{\varepsilon}^{12\:\dagger} & = & \bbox{\varepsilon}^{21}
\end{eqnarray}
for the noise. The stochastics properties of $\bbox{\varepsilon}$ could be 
derived from a microscopic Hamiltonian \cite{Dorokhov83a,Endesfelder96a}. 
Here, we adopt a simple model \cite{Mello92a,Chalker93a} which assumes 
Gaussian white noise such that 
\begin{eqnarray}
\label{ScatMod}
\langle\varepsilon_{kl}(x)\rangle & = & 0, \nonumber\\
\langle\varepsilon^{11}_{kl}(x)\varepsilon^{11\: *}_{k'l'}(x') \rangle & = & 
\frac{1}{l^{f}_{kl}}\delta_{kk'}\delta_{ll'}\delta(x-x'), \nonumber \\
\langle \varepsilon^{22}_{kl}(x)\varepsilon^{22\: *}_{k'l'}(x') \rangle & = & 
\frac{1}{l'^{f}_{\: kl}}\delta_{kk'}\delta_{ll'}\delta(x-x'), \nonumber \\
\langle \varepsilon^{12}_{kl}(x)\varepsilon^{12\: *}_{k'l'}(x') \rangle & = & 
\frac{1}{l^{b}_{kl}}\delta_{kk'}\delta_{ll'}\delta(x-x'),  
\end{eqnarray}
and all other independent second moments are zero. The mean free paths 
$l^{f}_{kl}$, $l'^{f}_{\: kl}$, and $l^{b}_{ij}$, $l'^{b}_{\: kl}$ for 
forward and backward scattering, respectively, are defined by the limits 
of the disorder averages 
\begin{eqnarray}
\frac{1}{l^{f}_{kl}} & \equiv & \lim_{\delta L \to 0}\frac{\langle 
                                |t_{kl}-\delta_{kl}|^{2}\rangle_{\delta L}}
                                {\delta L},
\nonumber\\
\frac{1}{l'^{f}_{\: kl}} & \equiv & \lim_{\delta L \to 0}\frac{\langle 
                                    |t'_{kl}-\delta_{kl}|^{2}\rangle_{
                                    \delta L}}{\delta L},
\nonumber\\
\frac{1}{l^{b}_{kl}} & \equiv & \lim_{\delta L \to 0}\frac{\langle 
                                |r_{kl}|^{2}\rangle_{\delta L}}{\delta 
                                L},
\nonumber\\
\frac{1}{l'^{b}_{\: kl}} & \equiv & \lim_{\delta L \to 0}\frac{\langle 
                                    |r'_{kl}|^{2}\rangle_{\delta L}}{
                                    \delta L}
\end{eqnarray}
for a short piece of conductor with length $\delta L$. Note that the 
symmetries (\ref{Symm2}) imply the relation $l^{b}_{kl}=l'^{b}_{\: lk}$. 

We want a path integral representation of the stochastic process 
(\ref{Lvgl1}) in terms of the transfer matrix $\bbox{T}$. The derivation 
technique which is most suited for that purpose derives the path integral 
directly from the Langevin equation (see chapter 4 in Ref. 
\onlinecite{Zinn-Justin93}). The symmetry constraints (\ref{Symm1}) on 
$\bbox{T}$ will be taken into account by $\delta$-functions which leads 
naturally to the invariant measure of the transfer matrix group as the 
path integration measure. We illustrate the essential ideas of the 
derivation technique with the simple example of diffusion on a circle 
before we deal with the transfer  matrix.             
\section{Diffusion on the circle as a simple example}
Let the angle $\varphi$ determine the position on a circle. The analogue
of the Langevin equation (\ref{Lvgl1}) is 
\begin{equation}
\label{Lvgl2}
\dot{u}\equiv \frac{du(t)}{dt}=\varepsilon(t)u(t)
\end{equation}   
where $u=\exp(i\varphi)$. The symmetry $\varepsilon^{*}=-\varepsilon$ 
implies $d(uu^{*})/dt=0$ which ensures that $u$ remains a 
phase. Choosing Gaussian white noise for the imaginary part of 
$\varepsilon$ such that 
\begin{eqnarray}
\langle\varepsilon(t)\rangle & = & 0, \nonumber\\
\langle\varepsilon(t)\varepsilon(t')^{*}\rangle & = & 2 D \delta(t-t')
\end{eqnarray}
leads to the FP equation
\begin{equation}
\frac{\partial p(t;\varphi)}{\partial t}=D 
\frac{\partial^{2} p(t;\varphi)}{(\partial\varphi)^{2}}
\end{equation}
which describes diffusion on the circle.

The probability distribution of $u$ can be formally expressed as 
\begin{equation}
p(t;u)=\langle\delta(u-\bar{u}(t))\rangle
\end{equation}
where $u\equiv u^{(1)}+iu^{(2)}$, $\delta(u)\equiv\delta(u^{(1)}) 
\delta(u^{(2)})$, and $\bar{u}(t)$ is the value of $u$ which is acquired 
at time $t$ for a certain realization of the noise and the initial value 
$\bar{u}(0)=u_{0}$. The brackets $\langle\cdots\rangle$ denote the 
average over all possible noise configurations. The path integral 
representation is derived by inserting a product of $\delta$-functions 
\begin{equation}
\label{pu2}
p(t;u)=\left\langle\int\prod_{t'=0}^{t} du(t') \delta(u(t')-\bar{u}(t'))
\delta(u(t)-u)\right\rangle
\end{equation}
where $du\equiv du^{(1)}du^{(2)}$. The $\delta$-function 
$\delta(u(t')-\bar{u}(t'))$ restricts the value of $u(t')$ to $\bar{u}(t')$. 
Since $\bar{u}(t')$ is not explicitly known we enforce this constraint
implicitly by the relation $\dot{u}(t)u^{-1}(t)-\varepsilon(t)=0$ 
which follows from the Langevin equation (\ref{Lvgl2}). That leads to 
\begin{eqnarray}
p(t;u) & = & \biggl\langle\int\prod_{t'=0}^{t} du(t') |\det \hat{\cal A}| 
\delta(\dot{u}(t')u^{-1}(t')-\varepsilon(t'))\nonumber\\
&  & \hspace{4cm}\times\delta(u(t)-u)\biggr\rangle
\end{eqnarray}
where the operator $\hat{\cal A}$ is defined by the functional derivative 
\begin{equation}
\label{Jacobian1}
{\cal A}_{jj'}(t,t') =  \frac{\delta (\dot{u}(t)u^{-1}(t)-
\varepsilon(t))^{(j)}} {\delta u^{(j')}(t')}.  
\end{equation}
The average over the Gaussian probability measure 
\begin{equation}
P[\varepsilon]\prod_{x=0}^{L}d\varepsilon(x)=\frac{1}{\cal N}\exp\left
\{-\int_{0}^{L}dx\frac{\varepsilon(x)\varepsilon^{*}(x)}{4D}\right\}
\prod_{x=0}^{L}d\varepsilon(x),
\end{equation}
where
\begin{equation}
d\varepsilon=d\varepsilon^{(1)}d\varepsilon^{(2)}\delta(\varepsilon+ 
\varepsilon^{*}),
\end{equation}
yields
\begin{eqnarray}
p(t;u) & = & \frac{1}{\cal N}\int\prod_{t'=0}^{t} du(t') \delta(\dot{u}
(t')u^{-1}(t')+\dot{u}^{*}(t')\nonumber\\
&  & \hspace*{1cm}\times u^{-1\:*}(t'))|\det \hat{\cal A}|\exp\{-S\},
\end{eqnarray}
where 
\begin{equation}
\label{S1}
S=\frac{1}{4D}\int_{0}^{t}dt'\dot{u}(t')u^{-1}(t')(\dot{u}(t') 
  u^{-1}(t'))^{*}
\end{equation}
and the path summation includes all paths which start at $u_{0}$ and
end at $u$. 

The property that $\dot{w}(t)w^{-1}(t)+\dot{w}^{*}(t)w^{-1\:*}(t)=
\dot{u}(t)u^{-1}(t)+\dot{u}^{*}(t)u^{-1\:*}(t)$ if $w(t)=u(t)v(t)$ and
$v(t)$ is a phase, suggests that $\prod_{t'=0}^{t} du(t') \delta(
\dot{u}(t')u^{-1}(t')+\dot{u}^{*}(t')u^{-1\:*}(t'))$ is proportional 
to $\prod_{t'=0}^{t} d\mu(u(t'))$ where $d\mu(u)$ is the invariant
measure on $U(1)$. This becomes explicit if the $\delta$-function is 
introduced via an auxiliary field $\kappa(t')$
\begin{equation}
p(t;u)=\int\prod_{t'=0}^{t} du(t')d\kappa(t')|\det \hat{\cal A}| 
       \exp\{-\tilde{S}\},
\end{equation}
where
\begin{eqnarray}
\tilde{S} & = & S+i\int_{0}^{t}dt'\kappa(t')(\dot{u}(t')u^{-1}(t')+
                \dot{u}^{*}(t')u^{-1\:*}(t'))\nonumber\\
          & = & S+i\int_{0}^{t}dt'\kappa(t')\frac{d}{dt'}\ln(u(t')
                u^{*}(t')).
\end{eqnarray}
Partial integration yields
\begin{equation}
\tilde{S}=S+i\int_{0}^{t}dt' \lambda(t')\ln(u(t')u^{*}(t')),
\end{equation}
where $\lambda(t)=-\dot{\kappa}(t)$. The Jacobian of the transformation
$\lambda(t)=-\dot{\kappa}(t)$ is an irrelevant constant which can be 
incorporated into the normalization factor. Hence
\begin{equation}
p(t;u)={\cal N}^{-1}\int\prod_{t'=0}^{t} d\mu(u(t'))|\det \hat{\cal A}| 
       \exp\{-S\}
\end{equation}
since $du\delta(\ln(uu^{*}))=du\delta(uu^{*}-1)$ which is proportional to
the invariant measure $d\mu(u)$ \cite{Endesfelder96a}. The restriction
to $uu^{*}=1$ in the invariant measure simplifies the action (\ref{S1}), 
\begin{equation}
S=\frac{1}{4D}\int_{0}^{t}dt'\dot{u}(t')\dot{u}^{*}(t'). 
\end{equation}
To calculate $\det{\cal A}$ we evaluate Eq. (\ref{Jacobian1}) which
gives 
\begin{eqnarray}
{\cal A}_{11}(t,t')& = & (a(t,t')+a^{*}(t,t'))/2, 
\nonumber\\
{\cal A}_{12}(t,t') & = & i(a(t,t')-a^{*}(t,t'))/2, 
\nonumber\\
{\cal A}_{21}(t,t') & = & -i(a(t,t')-a^{*}(t,t'))/2, 
\nonumber\\
{\cal A}_{22}(t,t') & = & (a(t,t')+a^{*}(t,t'))/2,   
\end{eqnarray}
where 
\begin{equation}
a(t,t')=u^{-1}(t)\left(\frac{d}{dt}\delta(t-t')-\delta(t-t')\dot{u}(t)
        u^{-1}(t)\right).
\end{equation}
The decomposition  $\hat{\cal A}=\hat{\cal B}\hat{\cal C}\hat{\cal D}$
into a product of three operators 
\begin{eqnarray}
[\hat{\cal B}](t,t') & = & \frac{1}{\sqrt{2}}
\left(\begin{array}{cc} \delta(t-t') & -i \delta(t-t') \\
                        -i \delta(t-t') & \delta(t-t') 
      \end{array}\right),\nonumber\\[1ex]
[\hat{\cal C}](t,t') & = & 
\left(\begin{array}{cc} a(t,t') & 0 \\
                        0 & a^{*}(t,t') 
      \end{array}\right),\nonumber\\[1ex]
[\hat{\cal D}](t,t') & = & \frac{1}{\sqrt{2}}
\left(\begin{array}{cc} \delta(t-t') & i \delta(t-t') \\
                        i \delta(t-t') & \delta(t-t') 
      \end{array}\right)    
\end{eqnarray}
implies that $\det\hat{\cal A}=\det\hat{C}=\det\hat{a}\det\hat{a}^{*}$
since $\det\hat{\cal B}=\det\hat{\cal D}=1$. The operator $\hat{a}$
can be as well factorized into $\hat{a}=\hat{a}_{1}\hat{a}_{2}\hat{a}_{3}$ 
where
\begin{eqnarray}
a_{1}(t,t') & = & u^{-1}(t)\delta(t-t') \nonumber\\
a_{2}(t,t') & = & \frac{d}{dt}\delta(t-t') \nonumber\\
a_{3}(t,t') & = & \delta(t-t')-\theta(t-t')\dot{u}(t')u^{-1}(t').
\end{eqnarray}
The determinant of $\hat{a}_{1}\hat{a}_{1}^{*}$ is one since the
$\delta$-function in the path integration measure enforces that $u(t)
u^{*}(t)=1$. The determinant of $\hat{a}_{2}$ is an irrelevant constant 
which contributes only to the normalization. Using $\det=\exp\mbox{tr}
\ln$ and $\ln(1+x)=\sum_{k=1}^{\infty}(-1)^{k+1}x^{k}/k$ to evaluate 
$\det\hat{a}_{3}$ yields
\begin{equation}
\det\hat{a}_{3}=\exp\left\{-\int_{0}^{t}dt' \theta(0)\dot{u}(t')u^{-1}(t')
                +\ldots\right\} 
\end{equation}
The higher order terms which are indicated by the dots vanish due to
products of $\theta$-functions. The quantity $\theta(0)$ is not defined 
yet which can be traced back to the multiplicative noise in the Langevin 
equation (\ref{Lvgl2}). The correct choice is $\theta(0)=1/2$ (see the
discussion in chapter 4 of Ref. \onlinecite{Zinn-Justin93}). Here this
choice does not matter since $\dot{u}(t')u^{-1}(t')+\dot{u}^{*}(t')u^{-1
\:*}(t') =0$ which implies that $\det\hat{a}_{3}\det\hat{a}_{3}^{*}=1$, 
leading to the final form
\begin{equation}
p(t;u)={\cal N}^{-1}\int\prod_{t'=0}^{t} d\mu(u(t'))\exp\{-S\}
\end{equation}
of the path integral representation of the stochastic process (\ref{Lvgl2}).
\section{The path integral for the transfer matrix}
\label{SPatIntT}
The analogue of Eq. (\ref{pu2}) for the transfer matrix is
\begin{eqnarray}
p(L;\bbox{T}) & = & \int\biggl\langle\int\prod_{x=0}^{L} d\bbox{T}(x)
                    \delta(\bbox{T}(x)-\bar{\bbox{T}}(x))\nonumber\\
              &   & \hspace{3cm}\times\delta(\bbox{T}(L)-\bbox{T})
                    \biggr\rangle,
\end{eqnarray}
where 
\begin{eqnarray}
d\bbox{T} & \equiv & \prod_{k,l} dT^{(1)}_{kl}dT^{(2)}_{kl} \nonumber\\
\delta(\bbox{T}-\bar{\bbox{T}})
& \equiv & \prod_{k,l} \delta(T^{(1)}_{kl}-\bar{T}^{(1)}_{kl})\nonumber\\
&        & \hspace{1cm} \times \delta(T^{(2)}_{kl}-\bar{T}^{(2)}_{kl}).
           \nonumber\\
& & 
\end{eqnarray}
Enforcing $\bar{\bbox{T}}(x)$ by $\dot{\bbox{T}}(x)\bbox{T}^{-1}(x)-
\bbox{\varepsilon}(x)=0$ which follows from the Langevin equation 
(\ref{Lvgl1}) yields 
\begin{eqnarray}
p(L;\bbox{T}) & = & \int\biggl\langle\int\prod_{x=0}^{L} dT(x)
                    |\det{\cal A}| \delta(\dot{\bbox{T}}(x)\bbox{T}^{-1}
                    (x)-\bbox{\varepsilon}(x)) \nonumber\\
              &   & \hspace{3cm}\times\delta(\bbox{T}(L)-\bbox{T})
                    \biggr\rangle,
\end{eqnarray}
where the operator $\hat{\cal A}$ is defined by the functional derivative
\begin{equation}
\label{Jacobian2}
{\cal A}^{jj'}_{kl,k'l'}(x,x') = \frac{\delta[\dot{\bbox{T}}(x)\bbox{T
}^{-1}(x)-\bbox{\varepsilon}(x)]^{(j)}_{kl}}{\delta T^{(j')}_{k'l'}(x')}.  
\end{equation}
Performing the average over the Gaussian probability measure 
\begin{eqnarray}
P[\bbox{\varepsilon}]\prod_{x=0}^{L}d\bbox{\varepsilon}(x) & = & \frac{1}
{\cal N}\exp\biggl\{-\frac{1}{2}\int_{0}^{L}dx\Bigl(l^{f}_{kl}\bbox{
\varepsilon}^{11}_{kl}(x)\bbox{\varepsilon}^{11\: *}_{kl}(x)\nonumber\\
&  & +l'^{f}_{\: kl}\bbox{\varepsilon}^{22}_{kl}(x)\bbox{\varepsilon
}^{22\:*}_{kl}(x)+l^{b}_{kl}\bbox{\varepsilon}^{12}_{kl}(x)
\bbox{\varepsilon}^{12\: *}_{kl}(x)
\nonumber\\
&  & +l'^{b}_{\: kl}\bbox{\varepsilon}^{21}_{kl}(x)\bbox{\varepsilon}^{21
\:*}_{kl}(x)\Bigr)\biggr\}\prod_{x=0}^{L}d\bbox{\varepsilon}(x),
\nonumber\\
\end{eqnarray}
where
\begin{eqnarray}
d\bbox{\varepsilon} & \equiv & \prod_{i,j,k,l} d\varepsilon^{ij\:(1)}_{kl}
                      \varepsilon^{ij\:(2)}_{kl}\delta_{\text{S}}(\bbox{
                      \varepsilon}), \nonumber\\
\delta_{\text{S}}(\bbox{\varepsilon})
& \equiv & \prod_{k<l} \Bigl\{\delta\Bigl(\bigl(\varepsilon^{11}_{kl}+
           \varepsilon^{11\: *}_{lk}\bigr)^{(1)}\Bigr)\delta\Bigl(\bigl(
           \varepsilon^{11}_{kl}+\varepsilon^{11\: *}_{lk}\bigr)^{(2)}
           \Bigr) \nonumber \\
&        & \hspace*{0.7cm}\times\delta\Bigl(\bigl(\varepsilon^{22}_{kl}+
           \varepsilon^{22\: *}_{lk}\bigr)^{(1)}\Bigr)\delta\Bigl(\bigl(
           \varepsilon^{22}_{kl}+\varepsilon^{22\: *}_{lk}\bigr)^{(2)}
           \Bigr)\Bigr\}\nonumber\\
&        & \prod_{k}\delta\Bigl(\bigl(\varepsilon^{11}_{kk}\bigr)^{(1)}
           \Bigr)\delta\Bigl(\bigl(\varepsilon^{22}_{kk}\bigr)^{(1)}
           \Bigr)\nonumber\\
&        & \prod_{k,l} \delta\Bigl(\bigl(\varepsilon^{12}_{kl}-
           \varepsilon^{21\: *}_{lk}\bigr)^{(1)}\Bigr)\delta\Bigl(\bigl(
           \varepsilon^{21\: *}_{lk}-\varepsilon^{12}_{kl}\bigr)^{(2)}
           \Bigr),\nonumber\\
\end{eqnarray}
yields
\begin{eqnarray}
p(L;\bbox{T}) & = & {\cal N}^{-1}\int\prod_{x=0}^{L} d\bbox{T}(x) 
                    \delta_{S}\Bigl(\dot{\bbox{T}}(x)\bbox{T}^{-1}(x)
                    \Bigr)\nonumber\\ 
              &   & \hspace{3.5cm}\times |\det\hat{\cal A}|\exp\{-S\},
\nonumber\\
\end{eqnarray}
where
\begin{eqnarray}
\label{S2}
S & = & \frac{1}{2}\int_{0}^{L}dx\Bigl\{l^{f}_{kl}[\dot{\bbox{T}}
\bbox{T}^{-1}]^{11}_{kl}[\dot{\bbox{T}}\bbox{T}^{-1}]^{11\: *}_{kl
}+l'^{f}_{\: kl}[\dot{\bbox{T}}\bbox{T}^{-1}]^{22}_{kl}\nonumber\\
&  & \hspace{1.5cm}\times[\dot{\bbox{T}}\bbox{T}^{-1}]^{22\:*}_{kl}+
l^{b}_{kl}[\dot{\bbox{T}}\bbox{T}^{-1}]^{12}_{kl}[\dot{\bbox{T}}
\bbox{T}^{-1}]^{12\: *}_{kl}\nonumber\\
&  & \hspace{1.5cm}+l'^{b}_{\: kl}[\dot{\bbox{T}}\bbox{T}^{-1}]^{21}_{kl}
[\dot{\bbox{T}}\bbox{T}^{-1}]^{21\:*}_{kl}\Bigr\}.\nonumber\\
\end{eqnarray}
By analogy with the previous section we expect that $\prod_{x=0}^{L} 
d\bbox{T}(x) \delta_{S}(\dot{\bbox{T}}(x)\bbox{T}^{-1}(x))$ is 
proportional to $\prod_{x=0}^{L}d\mu(\bbox{T}(x))$ where $d\mu(\bbox{T})$ 
is the invariant measure of the transfer matrix group. This will be 
proven in appendix \ref{InMeaT}. The form of the invariant measure in
terms of the polar coordinates (\ref{PolDecomp}) is
\begin{eqnarray}
d\mu(\bbox{T}) & = & \prod_{k<l}(\cosh\Gamma_{k}-\cosh\Gamma_{l})^{2}
\prod_{k}\sinh\Gamma_{k}d\Gamma_{k}\nonumber\\
&  & \times\prod_{k=1}^{4}d\mu(\bbox{u}_{k}),
\end{eqnarray}
where $d\mu(\bbox{u}_{k})$ is the the invariant measure on the unitary 
group \cite{Mello91b}.

We proceed with the calculation of $\det {\cal A}$. Using $\partial/
\partial T^{(1)}_{kl}=\partial/\partial T_{kl}+\partial/\partial 
T^{*}_{kl}$, $\partial/\partial T^{(2)}_{kl}=i(\partial/\partial T_{kl}
-\partial/\partial T^{*}_{kl})$, and $\partial T^{-1}_{kl}/ \partial 
T_{k'l'}=-T^{-1}_{kk'}T^{-1}_{l'l}$ to evaluate Eq. 
(\ref{Jacobian2}) yields 
\begin{eqnarray}
[\hat{\cal A}]^{11} & = & (\hat{\bbox{A}}+\hat{\bbox{A}}^{*})/2, 
\nonumber\\[1ex]
[\hat{\cal A}]^{12} & = & i(\hat{\bbox{A}}-\hat{\bbox{A}}^{*})/2, 
\nonumber\\[1ex]
[\hat{\cal A}]^{21} & = & -i(\hat{\bbox{A}}-\hat{\bbox{A}}^{*})/2, 
\nonumber\\[1ex]
[\hat{\cal A}]^{22} & = & (\hat{\bbox{A}}+\hat{\bbox{A}}^{*})/2,   
\end{eqnarray}
where 
\begin{eqnarray}
A_{kl,k'l'}(x,x') & = & \delta_{km}T^{-1}_{nl}(x)\biggl(\frac{d}{dx}
\delta(x-x')\delta_{mk'}\delta_{nl'}-\nonumber\\
&  & \delta(x-x')[\dot{\bbox{T}}(x)\bbox{T}^{-1}(x)]_{mk'}\delta_{nl'}
\biggr).
\end{eqnarray}
The decomposition  $\hat{\cal A}=\hat{\cal B}\hat{\cal C}\hat{\cal D}$
into a product of three operators 
\begin{eqnarray}
\hat{\cal B} & = & \frac{1}{\sqrt{2}}
                   \left(\begin{array}{cc} 
                   \hat{\bbox{1}} & -i\hat{\bbox{1}} \\
                   -i\hat{\bbox{1}} & \hat{\bbox{1}} 
                   \end{array}\right),\nonumber\\[1ex]
\hat{\cal C} & = & \left(\begin{array}{cc} 
                         \bbox{\hat{A}} & 0 \\
                         0 & \hat{\bbox{A}}^{*} 
                         \end{array}\right),\nonumber\\[1ex]
\hat{\cal D} & = & \frac{1}{\sqrt{2}}
                   \left(\begin{array}{cc} 
                   \hat{\bbox{1}}& i\hat{\bbox{1}} \\
                   i\hat{\bbox{1}} & \hat{\bbox{1}}
                   \end{array}\right),
\end{eqnarray}
where $[\hat{\bbox{1}}]_{kl,k'l'}(x,x')=\delta(x-x')\delta_{kk'}
\delta_{ll'}$, implies that $\det\hat{\cal A}=\det\hat{C}=\det\hat{
\bbox{A}}\det\hat{\bbox{A}}^{*}$ since $\det\hat{\cal B}=\det\hat{\cal D}
=1$. The operator $\hat{\bbox{A}}$ can be as well factorized into $\hat{ 
\bbox{A}}=\hat{\bbox{A}}_{1}\hat{\bbox{A}}_{2}\hat{\bbox{A}}_{3}$ 
where
\begin{eqnarray}
A_{1;kl,k'l'}(x,x') & = & [\bbox{1}\otimes(\bbox{T}^{-1})^{T}(x)]_{kk', 
ll'}\delta(x-x') \nonumber\\
A_{2;kl,k'l'}(x,x') & = & \frac{d}{dx}\delta(x-x')\delta_{kk'} 
\delta_{ll'} \nonumber\\
A_{3;kl,k'l'}(x,x') & = & [\delta(x-x')\bbox{1}\otimes\bbox{1}-
\theta(x-x')\nonumber\\
&  & \times\dot{\bbox{T}}(x')\bbox{T}^{-1}(x')\otimes\bbox{1}]_{kl,k'l'}.
\end{eqnarray}
The product $\det\hat{\bbox{A}}_{1}\det\hat{\bbox{A}}^{*}_{1}$ is one 
since the determinant of the transfer matrix is a phase. The determinant of 
$\hat{\bbox{A}}_{2}$ is an irrelevant constant which contributes only to 
the normalization. Using $\det=\exp\mbox{tr}\ln$ and $\ln(1+x)=\sum_{k=1
}^{\infty}(-1)^{k+1}x^{k}/k$ to evaluate $\det\hat{\bbox{A}}_{3}$ yields
\begin{equation}
\det\hat{\bbox{A}}_{3}=\exp\left\{-N\theta(0)\int_{0}^{L}dx\:\mbox{tr}(
                       \dot{\bbox{T}}(x)\bbox{T}^{-1}(x))\right\} 
\end{equation}
The symmetries of the transfer matrix imply that $\mbox{tr}(\dot{\bbox{T}
}(x)\bbox{T}^{-1}(x)+(\dot{\bbox{T}}(x)$$\bbox{T}^{-1}(x))^{*})=0$ which 
gives $\det\hat{\bbox{A}}_{3}\det\hat{\bbox{A}}_{3}^{*}=1$.

That leads to the final form    
\begin{equation}
p(L;\bbox{T}) = {\cal N}^{-1}\int\prod_{x=0}^{L} d\mu(\bbox{T}(x)) 
                \exp\{-S\}
\end{equation}
of the path integral, where $S$ is the action of Eq. (\ref{S2}).
\section{The DMPK equation}
We formulate the DMPK equation in terms of diffusion on the coset space 
of the transfer matrix group as has been done by H\"uffmann 
\cite{Hueffmann90a}. In our context that can be achieved with the 
equivalent channel model (ECM). This model has been introduced by Mello 
and Tomsovic for the orthogonal  universality class \cite{Mello91c,%
Mello92a}. They showed that it is equivalent to the DMPK equation with 
$\beta=1$, in the sense that the joint probability distributions for 
$\bbox{\Gamma}$ of both models are identical. The ECM for the unitary 
universality class is just the model (\ref{ScatMod}) with backscattering 
mean free paths of the form
\begin{equation}
\frac{1}{l^{b}_{mn}}=\frac{1}{l N}
\end{equation}
and arbitary forward scattering mean free paths. It is equivalent to the   
DMPK equation with $\beta=2$ in the same sense. The difference between 
the DMPK equation and the ECM is that the unitary matrices need not be 
isotropically distributed and that there can be correlations between 
them and $\bbox{\Gamma}$. 

We choose forward scattering to be infinitely strong so that the mean 
free paths $l^{f}_{mn}$ and $l'^{f}_{mn}$ are zero. Then, the action 
(\ref{S2}) simplifies 
\begin{eqnarray}
S & = & \frac{Nl}{2}\int_{0}^{L}dx\;\mbox{tr}\Bigl\{[\dot{\bbox{T}}
        \bbox{T}^{-1}]^{12}\Bigl([\dot{\bbox{T}}\bbox{T}^{-1}]^{12}
        \Bigr)^{\dagger} \nonumber\\
&  & \hspace*{3.5cm}+[\dot{\bbox{T}}\bbox{T}^{-1}]^{21}\left([\dot{
     \bbox{T}}\bbox{T}^{-1}]^{21}\right)^{\dagger}\Bigr\}.\nonumber\\
&  & 
\end{eqnarray}
Using that $\dot{\bbox{T}}\bbox{T}^{-1}=-\bbox{T}\dot{\bbox{T}^{-1}}$ 
and the symmetries of $\dot{\bbox{T}}\bbox{T}^{-1}$ one can simplify
further 
\begin{eqnarray}
S & = & \frac{Nl}{8}\int_{0}^{L}dx\;\mbox{tr}\Bigl\{\Bigr(
\dot{\bbox{T}}\bbox{T}^{-1}+\bigl(\dot{\bbox{T}}\bbox{T}^{-1}\bigr)^{
\dagger}\Bigr)^{2}\Bigr\}\nonumber\\
& = & \frac{Nl}{8}\int_{0}^{L}dx\;\mbox{tr}\Bigl\{2\;\dot{\bbox{T}}
\bbox{T}^{-1}\bigl(\dot{\bbox{T}}\bbox{T}^{-1}\bigr)^{\dagger}\nonumber\\
&   & \hspace*{4cm}-\dot{\bbox{T}}\dot{\bbox{T}^{-1}}-\dot{\bbox{T}}^{
\dagger}\dot{\bbox{T}^{-1}}^{\dagger}\Bigr\}\nonumber\\
& = & -\frac{Nl}{8}\int_{0}^{L}dx\;\mbox{tr}\Bigl(\dot{\bbox{M}}
\dot{\bbox{M}^{-1}}\Bigr),
\end{eqnarray}
where $\bbox{M}=\bbox{T}^{\dagger}\bbox{T}$ which does not depend on 
$\bbox{u}_{1}$ and $\bbox{u}_{3}$ anymore. The infinite strong forward 
scattering immediately randomizes the probability distribution of 
$\bbox{u}_{1}$ and $\bbox{u}_{3}$ so that they become isotropically 
distributed. Note that the space which is formed by the matrices 
$\bbox{M}$ is isomorphic to the coset space of the transfer matrix group.
The path integral describes diffusion on the coset space since the action 
is the classical action for free motion on this space 
\cite{Olshanetsky81a,Olshanetsky83a}. 

Introducing the dimensionless length $s=x/(Nl)$ yields
\begin{equation}
S = -\frac{1}{8}\int_{0}^{1/g_{cl}}ds\;\mbox{tr}\Bigl(\dot{\bbox{M}}
\dot{\bbox{M}^{-1}}\Bigr),
\end{equation}
where the dot now stands for the derivative with respect to $s$ and 
$g_{cl}\equiv Nl/L$ is the classical (bare) conductance \cite{Macedo94a,%
Macedo94b} in units of $e^{2}/h$. Hence, large conductances correspond to 
the 'short time' regime of the path integral which justifies a saddle point 
approach for good conductors. 
The variation $\bbox{M}(s)+\delta\bbox{M}(s)=\delta\bbox{T}^{\dagger}(s)
\bbox{M}(s)\delta\bbox{T}(s)$, where $\delta\bbox{T}=\bbox{1}+\bbox{
\varepsilon}$ and $\bbox{\varepsilon}$ obeys the symmetries (\ref{Symm2}) 
leads to the saddlepoint equation
\begin{eqnarray}
0 & = & \delta S\propto\int_{0}^{1/g_{cl}}ds\;\mbox{tr}\Bigl(\bigl(\bbox{
\varepsilon}^{\dagger}\bbox{M}+\bbox{M\varepsilon}\bigr)\ddot{\bbox{M}^{
-1}}\nonumber\\
& & \hspace{3cm} -\ddot{\bbox{M}}\bigl(\bbox{\varepsilon}\bbox{M}^{-1}
+\bbox{M}^{-1}\bbox{\varepsilon}^{\dagger}\bigr)\Bigr).
\end{eqnarray}
One can verify easily that $\bbox{M}_{sp}(s)=\exp\{s\bbox{X}\}$ is the
solution for a path which starts at $\bbox{M}(0)=\bbox{1}$ and ends at 
$\bbox{M}=\exp\{\bbox{X}/g_{cl}\}$. Evaluation of the saddle point action 
yields the transfer matrix probability measure in saddle point 
approximation
\begin{eqnarray}
p(L;\bbox{T})d\mu(\bbox{T}) & \approx & \prod_{k}\exp\left\{-\frac{Nl
}{4L}\Gamma^{2}_{k}\right\}d\mu(\bbox{T})\nonumber\\     
& = & \prod_{k<l}(\cosh\Gamma_{k}-\cosh\Gamma_{l})^{2}\prod_{k}\exp\left\{
-\frac{Nl}{4L}\Gamma^{2}_{k}\right\}\nonumber\\
&   & \times\prod_{k}\sinh\Gamma_{k}d\Gamma_{k}\prod_{k=1}^{4} d\mu(
\bbox{u}_{k}).
\end{eqnarray}
This is just the random matrix theory probability distribution measure 
which has been proposed for the transfer matrix \cite{Muttalib87a,%
Beenakker93a,Beenakker93b}. Since it is known that random transfer 
matrix theory describes the stochastic properties of ballistic conductors 
\cite{Baranger94a} we conclude that the saddle point of the path
integral correctly describes the ballistic regime of the conductor.
\section{Conclusion}
In summary we have presented a path integral approach to the stochastic 
properties of mesoscopic disordered conductors. Its application to 
quasi-one-dimensional wires in the ballistic regime led to the random 
transfer matrix theory probability distribution. We believe that known 
results for the quasi-one-dimensional wire could be recovered by a 
systematic perturbation expansion in powers of $1/g_{cl}$. At the 
moment it is not clear to us wether the 'short time regime' of the 
path integral in higher dimensions corresponds as well to conductors
with large conductances. That still has to be clarified. The further 
development of the path integral technique also remains to be done.  
\section*{Acknowledgments}
I would like to thank John Chalker for numerous useful discussions. I am
also indebted to Klaus Oerding for his help and encouragement in overcoming
technical difficulties. I have benefitted from discussions with Richard
Sza\-bo and the participants of the Cargese Summer School 1996 on Path 
Integrals during which part of this work has been completed. The research 
has been supported by the individual Human Capital and Mobility Grant No. 
ERBCHB1CT941365 of the European Union and by the Engineering and Physical 
Sciences Research Council (EPRSC) of Great Britain. 
\begin{appendix}
\section{The invariant measure of the transfer matrix group}
\label{InMeaT}
The invariant measure on the transfer matrix group does not change under 
multiplication with a fixed transfer matrix $\bbox{T}_{0}$ from the left 
or the right
\begin{equation}
\label{TinvM}
d\mu(\bbox{T})=d\mu(\bbox{T}_{0}\bbox{T})=d\mu(\bbox{T}\bbox{T}_{0}).
\end{equation}
In this appendix we prove the claim of sect. \ref{SPatIntT} that
$\prod_{x=0}^{L} d\bbox{T}(x) \delta_{S}\bigl(\dot{\bbox{T}}(x)\bbox{T
}^{-1}(x)\bigr)$ is proportional to $\prod_{x=0}^{L}d\mu(\bbox{T}(x))$.

Since the inverse of $\bbox{T}$ in $\delta_{S}$ cannot be handled as 
easily as $u^{-1}$ in the example of diffusion on the circle, we show 
first that $\delta_{S}(\bbox{\varepsilon})\propto\delta_{S}(\bbox{\Sigma
}_{z}\bbox{T}^{\dagger}\bbox{\Sigma}_{z}\bbox{\varepsilon}\bbox{T})$ up 
to a Jacobian. This will allow to replace $\dot{\bbox{T}}\bbox{T}^{-1}$ 
in the argument of $\delta_{S}$ by $\bbox{\Sigma}_{z}\bbox{T}^{\dagger}
\bbox{\Sigma}_{z}\dot{\bbox{T}}$. 

Writing the $\delta$-function in terms of its Fourier representation 
yields 
\begin{equation}
\delta_{\text{S}}(\bbox{\varepsilon})=\frac{1}
{(2\Pi)^{4N^{2}}}\int d\bbox{\kappa}\exp\biggl\{\frac{i}{2}\mbox{tr}
\bigl[\bbox{\kappa}\bigl(\bbox{\varepsilon}+\bbox{\Sigma}_{z}\bbox{
\varepsilon}^{\dagger}\bbox{\Sigma}_{z}\bigr)\bigr]\biggr\}
\end{equation}
where
\begin{equation}
\bbox{\kappa}=\left(\begin{array}{cc}
                     \bbox{\kappa}^{11} & \bbox{\kappa}^{12} \\
                     \bbox{\kappa}^{21} & \bbox{\kappa}^{22}
                     \end{array}\right),
\end{equation}
\begin{eqnarray}
\bbox{\kappa}^{11\:\dagger} & = & \bbox{\kappa}^{11}\nonumber\\
\bbox{\kappa}^{22\:\dagger} & = & \bbox{\kappa}^{22}\nonumber\\
\bbox{\kappa}^{12\:\dagger} & = & -\bbox{\kappa}^{21},
\end{eqnarray}
and
\begin{eqnarray}
d\bbox{\kappa} & = & \prod_{k<l} d\kappa^{11\:(1)}_{kl}d\kappa^{11\:(2)
}_{kl}d\kappa^{22\:(1)}_{kl}d\kappa^{22\:(2)}_{kl}\nonumber\\
&  & \prod_{k}d\kappa^{11\:(1)}_{kk}d\kappa^{22\:(1)}_{kk}
     \prod_{k,l}d\kappa^{12\:(1)}_{kl}d\kappa^{12\:(2)}_{kl}.
\end{eqnarray}
Then the linear transformation 
\begin{equation}
\label{linTr1}
\bbox{\varepsilon}'=\bbox{\Sigma}_{z}\bbox{T}^{\dagger}\bbox{\Sigma
}_{z}\bbox{\varepsilon}\bbox{T}
\end{equation}
of $\bbox{\varepsilon}$ can be absorbed into $\bbox{\kappa}$,
\begin{eqnarray}
\delta_{\text{S}}(\bbox{\varepsilon}') & = &
\frac{1}{(2\Pi)^{4N^{2}}}\int d\bbox{\kappa} \exp\biggl\{\frac{i}{2}
\mbox{tr}\bigl[\bbox{\kappa}'\bigl(\bbox{\varepsilon}+\bbox{\Sigma}_{z}
\bbox{\varepsilon}^{\dagger}\bbox{\Sigma}_{z}\bigr)\bigr]
\biggr\},\nonumber\\
&  &
\end{eqnarray}
where 
\begin{equation}
\label{linTr2}
\bbox{\kappa}'=\bbox{T}\bbox{\kappa}\bbox{\Sigma}_{z}
\bbox{T}^{\dagger}\bbox{\Sigma}_{z}.
\end{equation} 
Since $\bbox{\kappa}'$ has the same symmetries as $\kappa$ it follows 
that
\begin{equation}
\delta_{\text{S}}(\bbox{\varepsilon}')=\delta_{\text{S}}(\bbox{
\varepsilon})/|{\cal J}(\bbox{T})|,
\end{equation}
where ${\cal J}(\bbox{T})$ is the Jacobian of the linear transformation 
(\ref{linTr2}). Hence, replacement of the argument $\dot{\bbox{T}}\bbox{T
}^{-1}$ in $\delta_{\text S}$ by $\bbox{\Sigma}_{z}\bbox{T}^{\dagger}\bbox{
\Sigma}_{z}\dot{\bbox{T}}$ via the linear transformation (\ref{linTr1}) 
yields 
\begin{eqnarray}
\lefteqn{\prod_{x=0}^{L}\delta_{\text S}\Bigl(\dot{\bbox{T}}(x)\bbox{T
}^{-1}(x)\Bigr)\propto }
\hspace{1cm}\nonumber\\
&  & \int\prod_{x=0}^{L}d\bbox{\kappa}(x)|{\cal J}(\bbox{T}(x))|
\nonumber\\
&  & \hspace*{1cm}\times\exp\biggl\{i\int_{x=0}^{L}dx\mbox{tr}
\biggl[\bbox{\kappa}\frac{d}{dx}\bigl(\bbox{\Sigma}_{z}\bbox{T}^{
\dagger}\bbox{\Sigma}_{z}\bbox{T}\bigr)\biggr]\biggr\}.\nonumber\\
\end{eqnarray}
Partial integration and using that $\bbox{\Sigma}_{z}\bbox{T}^{\dagger}
\bbox{\Sigma}_{z}\bbox{T}=\bbox{1}$ at the endpoints gives
\begin{eqnarray}
\lefteqn{\prod_{x=0}^{L}\delta_{\text S}\Bigl(\dot{\bbox{T}}(x)\bbox{T
}^{-1}(x)\Bigr)\propto }
\hspace{1cm}\nonumber\\
&  & \int\prod_{x=0}^{L}d\bbox{\kappa}(x)|{\cal J}(\bbox{T}(x))|
\nonumber\\
&  & \hspace*{1cm}\times\exp\biggl\{-i\int_{x=0}^{L}dx\mbox{tr}
\bigl[\dot{\bbox{\kappa}}\bigl(\bbox{\Sigma}_{z}\bbox{T}^{\dagger}\bbox{
\Sigma}_{z}\bbox{T}-\bbox{1}\bigr)\bigr]\biggr\}.\nonumber\\
\end{eqnarray}
The Jacobian of the transformation $\tilde{\bbox{\kappa}}=-\dot{
\bbox{\kappa}}$ is a constant. Hence 
\begin{eqnarray}
\label{DeltaS1}
\lefteqn{\prod_{x=0}^{L}\delta_{\text S}\Bigl(\dot{\bbox{T}}(x)\bbox{T
}^{-1}(x)\Bigr)\propto }
\hspace{1cm}\nonumber\\
&  & \prod_{x=0}^{L}|{\cal J}(\bbox{T}(x))|\delta_{\text S}\bigl(\bbox{
\Sigma}_{z}\bbox{T}^{\dagger}(x)\bbox{\Sigma}_{z}\bbox{T}(x)-\bbox{1}
\bigr).\nonumber\\
\end{eqnarray}
In order to calculate ${\cal J}(\bbox{T})$ we introduce the $(4N^{2})$-
vector notation 
\begin{equation}
\vec{\kappa}^{T}=(\kappa_{11},\ldots,\kappa_{12N},\kappa_{21},\ldots,
\kappa_{2N2N})
\end{equation}
of the matrix $\bbox{\kappa}$. Then $\vec{\kappa}'=\left(\bbox{T}
\otimes(\bbox{\Sigma}_{z}\bbox{T}^{\dagger}\bbox{\Sigma}_{z})^{T}\right)
\vec{\kappa}$. There is a complex matrix $\bbox{E}$ such that $\vec{
\kappa}=\bbox{E}\vec{\kappa}_{ind}$, where $\vec{\kappa}_{\text ind}$ 
contains the $4N^{2}$ real and imaginary parts of the independent 
matrix elements of $\bbox{\kappa}$. Therefore
\begin{equation}
\vec{\kappa}'_{\text ind}=\bbox{E}^{-1}\left(\bbox{T}\otimes(\bbox{
\Sigma}_{z}\bbox{T}^{\dagger}\bbox{\Sigma}_{z})^{T}\right)\bbox{E}
\vec{\kappa}_{\text ind}.
\end{equation}
${\cal J}(\bbox{T})$ is the determinant of this linear transformation,
which is one since the $\delta$-functions in Eq. (\ref{DeltaS1}) enforces 
$\bbox{\Sigma}_{z}\bbox{T}^{\dagger}\bbox{\Sigma}_{z}$ to be the inverse 
of $\bbox{T}$. That leads to
\begin{equation}
\prod_{x=0}^{L}\delta_{\text S}\Bigl(\dot{\bbox{T}}(x)\bbox{T
}^{-1}(x)\Bigr)\propto \prod_{x=0}^{L}\delta_{\text S}\bigl(\bbox{
\Sigma}_{z}\bbox{T}^{\dagger}(x)\bbox{\Sigma}_{z}\bbox{T}(x)-\bbox{1}
\bigr).                               
\end{equation}
It remains to be shown that 
\begin{equation}
d\mu(\bbox{T})\equiv d\bbox{T}\delta_{\text S}\bigl(\bbox{\Sigma}_{z}
\bbox{T}^{\dagger}\bbox{\Sigma}_{z}\bbox{T}-\bbox{1}\bigr)
\end{equation}
has the properties (\ref{TinvM}) and therefore is the invariant measure.

For multiplication with a transfer matrix $\bbox{T}_{0}$ from the left the 
argument of the $\delta$-function does not change which leads to 
\begin{equation}
d\mu(\bbox{T}_{0}\bbox{T})=d\bbox{T}|{\cal I}(\bbox{T}_{0})|\delta_{
\text S}\bigl(\bbox{\Sigma}_{z}\bbox{T}^{\dagger}\bbox{\Sigma}_{z}\bbox{T}-
\bbox{1}\bigr),
\end{equation}
where ${\cal I}(\bbox{T}_{0})$ is the Jacobian of the linear 
transformation $\bbox{T}'=\bbox{T}_{0}\bbox{T}$. Expressing this 
transformation in terms of real vectors yields
\begin{equation}
\left(\begin{array}{c}
\vec{T}'^{(1)}\\
\vec{T}'^{(2)}
\end{array}\right)=
\left(\begin{array}{cc}
\bbox{T}_{0}^{(1)}\otimes\bbox{1} & - \bbox{T}_{0}^{(2)}\otimes\bbox{1}\\
\bbox{T}_{0}^{(2)}\otimes\bbox{1} & \bbox{T}_{0}^{(1)}\otimes\bbox{1} 
\end{array}\right)
\left(\begin{array}{c}
\vec{T}^{(1)}\\
\vec{T}^{(2)}
\end{array}\right).
\end{equation}
The Jacobian ${\cal I}(\bbox{T}_{0})$ is the determinant of the 
transformation matrix which can be decomposed into the product
\begin{equation}
\frac{1}{\sqrt{2}}\left(\begin{array}{cc}
\bbox{1} & -i\bbox{1}\\
-i\bbox{1} & \bbox{1} 
\end{array}\right)
\left(\begin{array}{cc}
\bbox{T}_{0} \otimes \bbox{1} & \bbox{0}\\
\bbox{0} & \bbox{T}_{0}^{*} \otimes \bbox{1} 
\end{array}\right)
\frac{1}{\sqrt{2}}\left(\begin{array}{cc}
\bbox{1} & i\bbox{1}\\
i\bbox{1} & \bbox{1} 
\end{array}\right)
\nonumber
\end{equation}
of three matrices. Since $\bbox{\Sigma}_{z}\bbox{T}_{0}^{\dagger}
\bbox{\Sigma}_{z}\bbox{T}_{0}=\bbox{1}$ implies that $\det\bbox{T}_{0}
\det\bbox{T}_{0}^{*}=1$ one finds that ${\cal I}(\bbox{T}_{0})=1$ and 
therefore $d\mu(\bbox{T}_{0}\bbox{T})=d\mu(\bbox{T})$.
 
Analogously it can be shown that the Jacobian for the multiplication 
with $\bbox{T}_{0}$ from the right is one as well which gives
\begin{equation}
d\mu(\bbox{T}\bbox{T}_{0})=d\bbox{T}\delta_{\text S}\bigl(\bbox{
\Sigma}_{z}\bbox{T}^{\dagger}_{0}\bbox{\Sigma}_{z}\bigl(\bbox{
\Sigma}_{z}\bbox{T}^{\dagger}\bbox{\Sigma}_{z}\bbox{T}-\bbox{1}\bigr)
\bbox{T}_{0}\bigr).
\end{equation}
As shown above $\delta_{\text S}\bigl(\bbox{\Sigma}_{z}\bbox{T}^{
\dagger}_{0}\bbox{\Sigma}_{z}\bbox{\varepsilon}\bbox{T}_{0}\bigr)=
\delta_{\text S}\bigl(\bbox{\varepsilon}\bigr)$. Hence 
\begin{eqnarray}
d\mu(\bbox{T}\bbox{T}_{0}) & = & d\bbox{T}\delta_{\text S}\bigl(\bbox{
\Sigma}_{z}\bbox{T}^{\dagger}\bbox{\Sigma}_{z}\bbox{T}-\bbox{1}\bigr)
\nonumber\\
& = & d\mu(\bbox{T})
\end{eqnarray}
which proves our claim.
\end{appendix}
 
\end{multicols}
\end{document}